# Luminosity Function of Early-Type Galaxies in Clusters Cores (as Selected by ANN)

Emilio Molinari

*Osservatorio Astronomico di Brera, via E. Bianchi 46, I-22055 Merate, LC, Italy*

Riccardo Smareglia

*Osservatorio Astronomico di Trieste, via Tiepolo 11, I-34131, Trieste, TS, Italy*

**Abstract.** We show the possibility to exploit the non-linear behaviour of artificial neural network (ANN) for the identification of the early-type component of the population of the core of clusters of galaxies. We result in a measurement of the luminosity function of the E/S0 galaxies which disfavours the hypothesis of a universal LF and calls for dynamical influences.

## 1. Introduction

The determination of the luminosity function (LF) of galaxies in clusters has now reached the key point of a true challenge to the supposed universality of its form (Colless 1989). The manifest presence of the two distinct classes of normal and dwarf galaxies and their different dynamical and luminosity evolution lead to the search of substantial differences from one LF to the other. Recent result obtained by Biviano et al. (1995), Lopez & Yee (1995) and Molinari (1995) evidence the peculiarities and the possible dependence on cluster parameters like richness and dynamical status of the overall shape of the observed luminosity function.

## 2. The Data-Base

Our data-base consists of a collection of photometric catalogs obtained from different observations at the 3.6m diameter ESO telescope (Molinari et al. 1989, 1994, 1996). Weather conditions and internal dome seeing did not allow to obtain concentrated images, and the overall seeing measurements did not fall below 1.6 arcsec, with a median value above 2.0 arcsec. This discarded completely the possibility to perform a visual-based classification of the objects in the fields, especially for the more distant clusters.

The whole process of classification will then be based on the search of structures in the abstract, 4-dimensional space of the $r$ magnitude, $g$-$r$ and $g$-$i$ colors, $R_{iso}$ isophotal radius values.

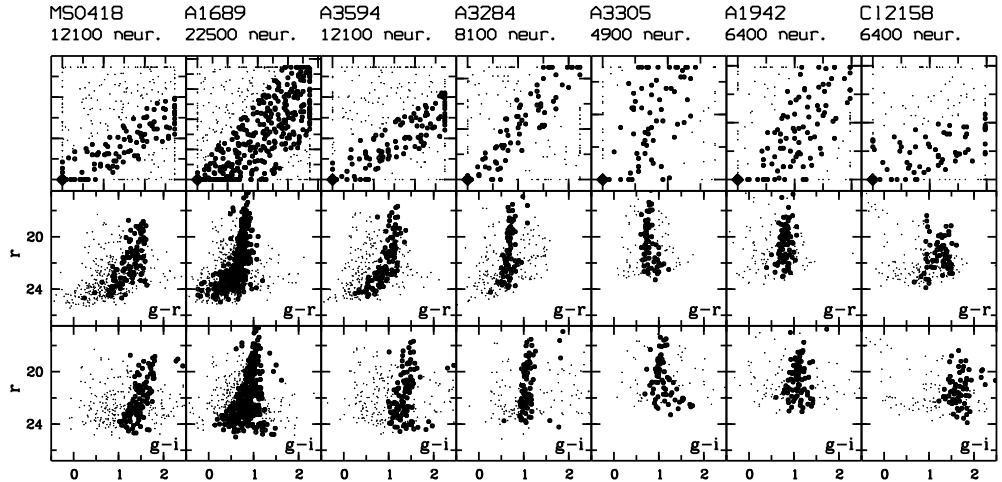

Figure 1. Identification of the E/S0 galaxies. In the upper panel for each cluster the SOM plane is shown, E/S0 are marked points while small dots show all the object in the catalog. The filled losange shows the BECM position. In the middle and lower panels the color-magnitude plots ($r$, $g$-$r$ and $r$, $g$-$i$) shows the early-type galaxies in the more usual planes.

## 3. The Self Organizing Map

The Self Organizing Map (SOM) algorithm developed by T. Kohonen (Kohonen 1984) creates a vector quantizer by adjusting weights from $N$ input nodes to $M$ output nodes arranged in a two dimensional grid, i.e. it defines a mapping from the input space $R^N$ ( in this case $N = 4$ ) onto a regular two-dimensional array of nodes. The process in which the SOM is formed is an unsupervised learning and it is used to find *clustering* of data in the input set (i.e. the photometric catalog) and to obtain a 2-D ordered map, where the *clustered* data are neatly arranged (see the top row in Figure 1). For each cluster we have created an output quadratic map which dimensions are proportional to the input data.

The direct human intervention is necessary to understand the relocation on the 2-D map of the interesting classes of which the input set is composed of. In our case it was easy to identify some fiducial early-type galaxies selecting a sample of objects that were falling well on the ridge of the color-magnitude ($c$-$m$) relation in the $r$ vs. $g$-$r$ plane, which is obeyed by all E/S0 galaxies.

Figure 1 shows the location of the E/S0 galaxies in the SOM plane, as well as in the two color-magnitude planes $r$ vs. $g$-$r$ and $r$ vs. $g$-$i$ in which it is easy to recognize the usual $c$-$m$ relation. These are the objects classified as E/S0 in our procedure. It has to be noticed the degenerated point (0,0) which includes probably a whole family of objects: there is the possibility that very bright spiral galaxies are included, due to the great relative importance of the $g$ (normalized) values which flattens a region of the 4-dimensional phase space in one single point. This warning applies to all the 7 clusters, but the influence on the luminosity functions can be neglected, as at most it is introducing 1–2 galaxies in the brightmost 0.5 mag bin. All the ANN selected objects will then be taken as fiducial early-type galaxies in the next section.

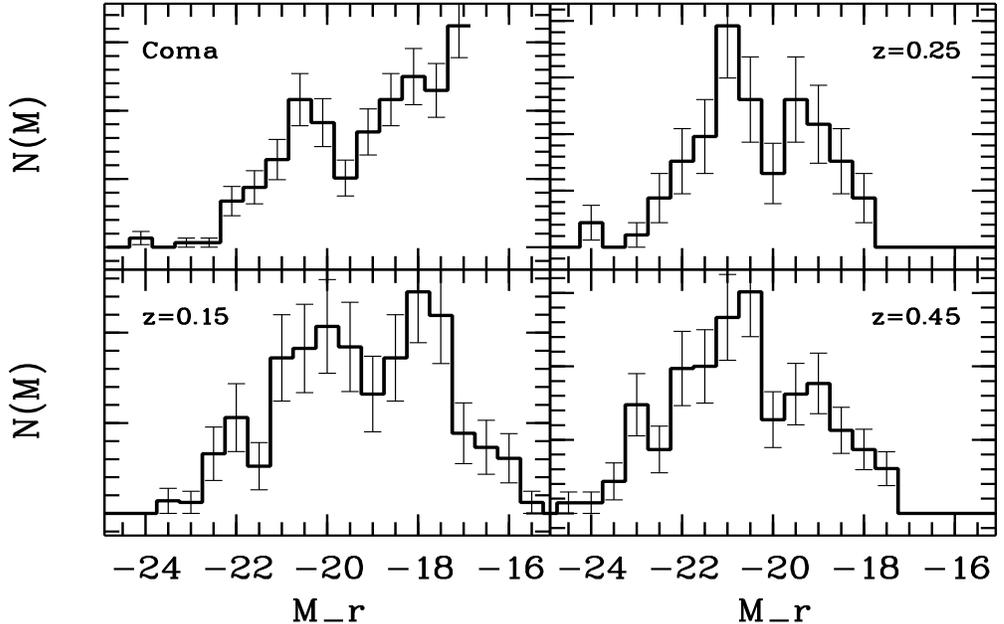

Figure 2. Luminosity function for 4 classes of clusters: the $z \sim 0$ Coma Cluster by Biviano et al. (1995); the $z \sim 0.15$, $z \sim 0.25$ and $z \sim 0.45$ group our 6 clusters.

## 4. The Luminosity Functions

We take now the assumption that the SOM machine is able to extract the member E/S0 galaxies from the photometric catalogs, and do not therefore apply any other correction for the background objects distribution. This is possible due to the fact that only the member galaxies obey the c-m relation an thus populate the 4-dimensional space in the relative hyper-region. The additional constraint of the two color-magnitude relations being satisfied at the same time gives a higher degree of confidence to our assumption.

Due to paucity of objects with complete $g$, $r$, $i$ photometry in some clusters we divided the whole set of clusters in three redshift intervals, which we label as 0.15, 0.25 and 0.45. In Fig. 3 the luminosity functions are shown. Each LF is computed summing the contribution for the cluster weigthed by their total number of galaxies. Error bars shown are computed accordingly.

The top panel of Figure 2 reports the LF for the Coma cluster (a well known virialized cluster) as determined by Biviano et al. (1995). Where Biviano et al. (1995) find a gap in the region of the change of regime (from normal to dwarfs E, $M_r \sim -19.5$) we also find significant lack of galaxies in the $z \sim 0.25$ interval.

The correspondent regions in the low redshift clusters is less significant, while in the high redshift ones falls just at the beginning of the faint incompleteness limit, showing nevertheless a breakdown at a more than $2\sigma$ level.

Interestingly, significant gaps in the magnitude distribution are present also at brighter magnitudes, near to the bright end of the LF of the low and high redshift clusters. The presence of cD galaxies can lead to depletion in the bright region of the LF, due to cannibalism (Hausman & Ostriker 1978) or tidal stripping through two-body interactions (Miller 1983). In a hierarchical scenario of cluster virialization depletion of the luminosity function could intervene at different magnitudes during the cluster dynamical evolution.

**Acknowledgments.** We are grateful to UIMP and the Organization for financial support and pleasant stay in Valencia, respectively.